\documentclass[%
 reprint,
superscriptaddress,
nofootinbib,
 amsmath,amssymb,
 aps,
]{revtex4-2}

\usepackage{graphicx}
\usepackage{dcolumn}
\usepackage{bm}
\usepackage{hyperref}

\newcommand{\BS}[1]{{\boldsymbol #1}}

\newcommand{\beq}{\begin{equation}}
\newcommand{\eeq}{\end{equation}}
\newcommand{\bea}{\begin{eqnarray}}
\newcommand{\eea}{\end{eqnarray}}

\newcommand{\half}{\frac 12}

\newcommand{\Slash}[1]{{\ooalign{\hfil#1\hfil\crcr\raise.167ex\hbox{/}}}}

\begin{document}

\title{The shape of quark flavors
}
\author{Shinsuke Kawai}
	\email{kawai@sci.kj.yamagata-u.ac.jp}
	\affiliation{%
    Faculty of Science, Yamagata University,
    1-4-12 Kojirakawa-machi, Yamagata, 990-8560 Japan
}%
\author{Nobuchika Okada}
 \email{okadan@ua.edu}
    \affiliation{%
    Department of Physics and Astronomy, University of Alabama, Tuscaloosa, Alabama, AL35487 USA
}%


\date{\today}

\begin{abstract}
We construct the Yukawa couplings of the quark sector as overlap integrals of Gaussian wave functions in extra spatial dimensions. 
Assuming that the wave function of each chiral fermion is localised at a point in the extra dimensions, while that of the Higgs doublet is flat, the Yukawa matrix elements are determined by the relative positions of the three left-handed quark doublets, $q_1$, $q_2$, $q_3$, and the six right-handed quark singlets, $u_1$, $u_2$, $u_3$, $d_1$, $d_2$, $d_3$.
We employ numerical optimisation techniques from machine learning to identify field configurations that satisfy all current experimental constraints on the quark Yukawa sector. 
In this framework, the large hierarchy of quark masses arises naturally, since the overlap integrals of Gaussian wave functions are exponentially sensitive to ${\mathcal O}(1)$ separations in the extra dimensions.
\end{abstract}

\maketitle


\section{\label{sec:intro}Introduction}

The flavour sector is often regarded as the least aesthetically appealing part of the Standard Model of particle physics, as it contains numerous free parameters that lack a compelling underlying explanation. 
This stands in contrast to the gauge sector, whose structure is dictated by the gauge principle. 
In addition, the fermion mass spectrum exhibits striking hierarchies; for example, the ratio of the top- and up-quark masses is as large as $m_t/m_u \sim 10^5$. 
This apparent arbitrariness of the flavour sector strongly suggests the existence of physics beyond the Standard Model, for which grand unified theories (GUTs) provide a natural framework. 
Reproducing the observed flavour parameters within GUTs remains a significant challenge: minimal GUT models are generally incompatible with experimental observations \cite{Georgi:1974sy}, whereas non-minimal extensions often suffer from a loss of predictive power due to their increased flexibility \cite{Georgi:1979df,Ellis:1979fg,Kawai:2024pws,Abu-Ajamieh:2025mjk}. 
Beyond the GUT framework, numerous bottom-up and phenomenological approaches have been proposed to address the flavour puzzle. 
Notable examples that have achieved partial success include the Froggatt-Nielsen mechanism \cite{Froggatt:1978nt} and the democratic mass matrix model \cite{Harari:1978yi}.

In this paper, we propose a new heuristic model of the flavour sector that is more {\em data-driven} than previous approaches. 
We assume that the Yukawa couplings arise from overlap integrals of Gaussian wave functions for chiral fermions localised in one and two extra dimensions, such that the Yukawa matrix elements are determined by the relative separations of the corresponding fermion fields in the extra-dimensional space. 
The construction of Yukawa couplings from wave-function overlaps in extra dimensions is not new and has been realised in scenarios such as split fermions \cite{Arkani-Hamed:1999ylh}, warped extra dimensions \cite{Grossman:1999ra}, magnetised extra dimensions \cite{Cremades:2004wa}, and F-theory GUT models \cite{Font:2009gq}. 
Our focus here is instead on reconstructing the observed quark masses and Cabibbo-Kobayashi-Maskawa (CKM) matrix parameters from the locations of wave packets in the extra dimensions; we consider cases with one and two extra dimensions that possess sufficient freedom to reproduce these observables. 
We restrict our attention to the quark sector in the present work, as the lepton sector is subject to additional ambiguities associated with the neutrino spectrum and will be discussed elsewhere. 
Determining fermion configurations in the extra dimensions that are consistent with experimental observations is technically challenging, as it requires solving a highly nonlinear problem in a high-dimensional parameter space. 
We address this challenge using numerical optimisation techniques commonly employed in machine learning. 
We show that this method yields fermion configurations that are consistent with current observations at the $\lesssim 1\sigma$ level. 
In particular, the observed mass hierarchy of the four-dimensional fermions can be naturally explained by the exponential dependence of the overlap integrals on ${\mathcal O}(1)$ separations in the extra dimensions.

The remainder of this paper is organised as follows. 
In the next section, we introduce our model of the quark sector. 
Section~\ref{sec:5d} presents the numerical methods used in our analysis and the corresponding results for the model with one extra dimension, while Sec.~\ref{sec:6d} is devoted to the case with two extra dimensions. 
We conclude in Sec.~\ref{sec:final} with brief remarks.

\section{\label{sec:xd_yukawa}Yukawa couplings as overlap integrals over extra dimensions}

\subsection{\label{sec:formalism}The formalism}

We start by assuming that the wave functions of the Standard Model chiral fermions are accompanied by components associated with $d$ extra dimensions, which are Gaussian with a common width.
For the left handed doublet quarks and right handed singlet quarks these are
\begin{align}
	q_L^i(\BS{\xi})=\left(\frac\pi 2\right)^{-d/4}
	e^{-(\BS{\xi}-\mathbf{q}_i)^2},\crcr
	u_R^i(\BS{\xi})=\left(\frac\pi 2\right)^{-d/4}
	e^{-(\BS{\xi}-\mathbf{u}_i)^2},\crcr
	d_R^i(\BS{\xi})=\left(\frac\pi 2\right)^{-d/4}
	e^{-(\BS{\xi}-\mathbf{d}_i)^2},	
\end{align}
where $\BS{\xi}$ is the coordinates of the $d$ extra dimensions, $\mathbf{q}_i$, $\mathbf{u}_i$, $\mathbf{d}_i$ are field-specific constant vectors (i.e. locations of these fields in the extra dimensions) and $i=1,2,3$. 
The Standard Model up- and down-type Yukawa matrices arise as effective couplings involving overlap integrals over the extra dimensions
\begin{align}
	y^u_{ij}
		=Y^u_{ij}\int d^d\xi\, q_L^i(\BS{\xi})u_R^j(\BS{\xi}) 
		=Y^u_{ij}e^{-\half(\mathbf{q}_i-\mathbf{u}_j)^2},\crcr
	y^d_{ij}
		=Y^d_{ij}\int d^d\xi\, q_L^i(\BS{\xi})d_R^j(\BS{\xi}) 
		=Y^d_{ij}e^{-\half(\mathbf{q}_i-\mathbf{d}_j)^2}.
\end{align}
The Yukawa components $Y^u_{ij}$, $Y^d_{ij}$ of the full $4+d$ dimensional theory are assumed to have unit modulus, $|Y^u_{ij}|=|Y^d_{ij}|=1$.
Some of their phases can be absorbed into unphysical phases of fermion wave functions; we give our parametrisation for specific cases of one and two extra dimensions below.

\subsection{\label{sec:5d_model}One extra dimension}

When there is only one extra dimension, the position of a field along the extra dimension is specified by a real number. 
One may choose a basis in which the phases of the up-type Yukawa couplings are trivial, such that $Y^u_{ij}=1$.
In this basis, however, not all phases in the down-type Yukawa matrix can be removed. 
We therefore allow non-trivial phases in $Y^d_{ij}$ for $i,j=2,3$.
Denoting 
\begin{align}
	f(\xi,\eta)\equiv e^{-\frac 12(\xi-\eta)^2},
\end{align} 
the 4-dimensional effective Yukawa couplings are thus parametrised as
\begin{align}\label{eqn:y5}
	y^u_{ij}=&f(q_i,u_j),\crcr
	y^d_{ij}=&\begin{pmatrix}
	f(q_1,d_1) & f(q_1,d_2) & f(q_1,d_3)\\
	f(q_2,d_1) & e^{ix_1}f(q_2,d_2) & e^{ix_2}f(q_2,d_3)\\
	f(q_3,d_1) & e^{ix_3}f(q_3,d_2) & e^{ix_4}f(q_3,d_3)
\end{pmatrix},
\end{align}
where $x_1$, $x_2$, $x_3$, $x_4$ parametrise the remaining phase degrees of freedom.
Since only the relative positions of the fields matter in this construction, we may set one of the field locations, $q_3$, to be fixed.
Then the field positions are parametrised by eight real numbers $x_5,\ldots,x_{12}$ as
\begin{alignat}{3}\label{eqn:coord5}
	q_1 &= x_5, &\quad q_2 &= x_6, &\quad q_3 &= 0,\crcr
	u_1 &= x_7, &\quad u_2 &= x_8, &\quad u_3 &= x_9,\crcr
	d_1 &= x_{10}, &\quad d_2 &= x_{11}, &\quad d_3 &= x_{12}.
\end{alignat}
Together with the phase parameters, the effective Yukawa couplings are thus controlled by 12 parameters $x_1$, $x_2,\cdots,x_{12}$, which will be adjusted to satisfy phenomenological constraints.

\begin{table*}[t]
\caption{\label{tab:table_5dparams}Optimised parameter values $x_1, x_2, \ldots, x_{12}$ for the 5 best fit samples for the one extra dimension model.
The optimised loss function values of these 5 samples are 0.942, 1.34, 1.37, 2.20, and 2.54.}
\begin{ruledtabular}
\begin{tabular}{ccccccccccccc}
 Sample &\multicolumn{12}{c}{Optimised parameter values}\\
 &$x_1$&$x_2$&$x_3$&$x_4$&$x_5$&$x_6$&$x_7$&$x_8$&$x_9$&$x_{10}$&$x_{11}$&$x_{12}$\\ \hline
\#1& -0.74222 & 1.3478 & 5.4744 & -4.4582 &  6.8517 & -0.022080 & 4.4409 & 6.3030 & -3.5100 & -4.5038 & 3.9302 & -4.1567 \\
\#2& -1.6859 & 1.6635 & -0.17823 & 0.20985 & 0.056992 & 6.8938 & 3.5297 & 6.3481 & -4.6365 & -4.2183 & 3.9720 & -4.1796 \\
\#3& -0.83353 & 1.4553 & -1.0119 & -4.6180 & 6.8937 & 0.056854 & 6.3481 & 3.5295 & -4.6370 & -4.2187 & 3.9719 & -4.1800 \\
\#4& 4.9318 & -0.38551 & -1.2008 & 4.2447 & -6.9569 & 0.55474 & -3.4003 & -6.4104 & -4.2314 & 5.2905 & -3.9216 & -3.9156 \\ 
\#5& 0.74028 & 0.87891 & 0.85937 & 0.31770 & -0.019286 & -7.0584 & -6.5114 & 3.4890 & -4.4305 & -3.9903 & 4.6432 & -4.0458 \\
\end{tabular}
\end{ruledtabular}
\end{table*}


\subsection{\label{sec:6d_model}Hermitian Yukawa from two extra dimensions}

While the strong CP problem is commonly addressed e.g. by the Peccei-Quinn mechanism \cite{Peccei:1977hh}, one may alternatively consider constructing a CP-free quark Yukawa sector from the outset\footnote{
In this case the topological CP phase needs to vanish separately.}.
One way of doing this is to assume Hermiticity of the Yukawa matrices, $(y^u)^\dag = y^u$ and $(y^d)^\dag = y^d$.
When there is only one extra dimension, the Hermiticity conditions are too restrictive to allow viable field configurations. 
This motivates us to consider constructing Hermitian effective Yukawa matrices using overlap integrals over two extra dimensions.

Assuming two extra dimensions $d=2$, again we can parametrise the up-type Yukawa to be phase-free, $Y^u_{ij}=1$, giving 
\begin{align}\label{eqn:yu6}
	y^u_{ij}=f(\mathbf{q}_i,\mathbf{u}_j),
\end{align}
where 
\begin{align}
	f(\BS{\xi},\BS{\eta})\equiv e^{-\half(\BS{\xi}-\BS{\eta})^2}
\end{align}
is the exponentiated separation in the two extra dimensions.
The remaining phases of the down-type Yukawa $Y^d_{ij}$ need to be consistent with Hermiticity, so we write
\begin{align}\label{eqn:yd6}
	y^d_{ij}=\begin{pmatrix}
f(\mathbf{q}_1,\mathbf{d}_1) & e^{ix_3}f(\mathbf{q}_1,\mathbf{d}_2) & e^{ix_2}f(\mathbf{q}_1,\mathbf{d}_3)\\
e^{-ix_3}f(\mathbf{q}_2,\mathbf{d}_1) & f(\mathbf{q}_2,\mathbf{d}_2) & e^{ix_1}f(\mathbf{q}_2,\mathbf{d}_3)\\
e^{-ix_2}f(\mathbf{q}_3,\mathbf{d}_1) & e^{-ix_1}f(\mathbf{q}_3,\mathbf{d}_2) & f(\mathbf{q}_3,\mathbf{d}_3),
\end{pmatrix}
\end{align}
with real parameters $x_1$, $x_2$, $x_3$.
The locations of the fields in the extra dimensions are parametrised as
\begin{alignat}{3}\label{eqn:coord6}
	\mathbf{q}_1 &= (0,0), &\quad \mathbf{q}_2 &= (0,x_4), &\quad \mathbf{q}_3 &= (x_5,x_6),\crcr
	\mathbf{u}_1 &= (z_1,w_1), &\quad \mathbf{u}_2 &= (z_2,w_2), &\quad \mathbf{u}_3 &= (z_3,w_3),\crcr
	\mathbf{d}_1 &= (\widetilde{z}_1,\widetilde{w}_1), &\quad \mathbf{d}_2 &= (\widetilde{z}_2,\widetilde{w}_2), &\quad \mathbf{d}_3 &= (\widetilde{z}_3,\widetilde{w}_3),
\end{alignat}
where $x_4$, $x_5$, $x_6$ are real parameters, and $z_i$, $w_i$, $\widetilde{z}_i$, $\widetilde{w}_i$ are subject to the Hermiticity conditions.

Let us first consider the Hermiticity conditions for the up-type Yukawa matrix.
Since its components are all real, $y^u = (y^u)^\dagger$ simply translates to relations between the off-diagonal elements, 
\begin{align}\label{eqn:HermiteConds}
	z_2^2+w_2^2 &= z_1^2+(x_4-w_1)^2,\crcr
	z_3^2+w_3^2 &= x_5-z_1)^2 + (x_6-w_1)^2,\crcr
	z_3^2 + (x_4-w_3)^2 &= (x_5-z_2)^2 + (x_6-w_2)^2.
\end{align}
Three of the six variables $z_1$, $z_2$, $z_3$, $w_1$, $w_2$, $w_3$ can be eliminated using the three relations above. 
Solving these relations for $z_3$, $w_2$ and $w_3$ we find,
\begin{align}\label{eqn:w2}
	w_2 &= \pm\sqrt{(w_1-x_4)^2 + z_1^2 - z_2^2},\\
	\label{eqn:w3}
	w_3 &= w_1+\frac{x_5(z_2-z_1)+x_6(w_2-w_1)}{x_4},\\
	\label{eqn:z3}
	z_3 &= \pm\sqrt{(x_5-z_1)^2+(x_6-w_1)^2-w_3^2},
\end{align}
where $w_3$ in the second line uses $w_2$ solved in the first line, and $z_3$ in the third line uses $w_3$ and $w_2$ solved in the first and the second line.  
The remaining three variables in the up-type Yukawa are $z_1$, $z_2$ and $w_1$; to them we assign parameters $x_7$, $x_8$, $x_9$ as
\begin{align}
	z_1=x_7,\quad 
	z_2=x_8,\quad 
	w_1=x_9.
\end{align}
The variables of the down-type Yukawa matrix can be treated similarly, as the phase factors of \eqref{eqn:yd6} are already chosen to be consistent with Hermiticity.
The off diagonal elements of $y^d = (y^d)^\dagger$ give three relations similar to \eqref{eqn:HermiteConds}, which are solved for $\widetilde z_3$, $\widetilde w_2$, and $\widetilde w_3$ as
\begin{align}\label{eqn:w2t}
	\widetilde{w}_2 &= \pm\sqrt{(\widetilde{w}_1-x_4)^2 + \widetilde{z}_1^2 - \widetilde{z}_2^2},\\
	\label{eqn:w3t}
	\widetilde{w}_3 &= \widetilde{w}_1+\frac{x_5(\widetilde{z}_2-\widetilde{z}_1)+x_6(\widetilde{w}_2-\widetilde{w}_1)}{x_4},\\
	\label{eqn:z3t}
	\widetilde{z}_3 &= \pm\sqrt{(x_5-\widetilde{z}_1)^2+(x_6-\widetilde{w}_1)^2-\widetilde{w}_3^2}.
\end{align}
Three variables $\tilde{z}_1$, $\tilde{z}_2$ and $\tilde{w}_1$ remain independent and we will assign parameters as
\begin{align}
	\widetilde z_1=x_{10},\quad 
	\widetilde z_2=x_{11},\quad 
	\widetilde w_1=x_{12}.
\end{align}
To summarise, the Hermitian Yukawa matrices in this two extra dimension construction are also controlled by 12 real parameters $x_1$, $x_2,\ldots,x_{12}$.

We note that the signs of $w_2$, $z_3$, $\widetilde{w}_2$, $\widetilde{z}_3$ in \eqref{eqn:w2}, \eqref{eqn:z3}, \eqref{eqn:w2t}, \eqref{eqn:z3t} can be chosen independently, resulting in $2^4 = 16$ possible sign configurations.
To distinguish them, we pick up the sign bit for each of $w_2$, $z_3$, $\widetilde{w}_2$, $\widetilde{z}_3$ (0 for $+$ and 1 for $-$), in this order, and label the corresponding solution type by the hexadecimal digit ($0,1,\ldots,9,a,b,\ldots,f$) representing the resulting 4-bit binary number.
For example, type $1$ corresponds to the sign configuration $(+,+,+,-)=(0001)$, type $e$ corresponds to $(-,-,-,+)=(1110)$, and so on.

\begin{table*}[t]
\caption{\label{tab:table_5dfit}
Predictions for the Standard Model parameters obtained from the five best-fit samples listed in Table~\ref{tab:table_5dparams}. 
The reference experimental values are the $\overline{\rm MS}$ quantities evaluated at 1 TeV \cite{Antusch:2025fpm}. 
The $1\sigma$ uncertainties are used to define the weights in the loss function.
}
\begin{ruledtabular}
\begin{tabular}{cccccccccccc}
 &\multicolumn{11}{c}{Standard Model parameters}\\
 &$y_u/10^{-6}$&$y_d/10^{-5}$&$y_s/10^{-4}$&$y_c/10^{-3}$&$y_b/10^{-2}$&$y_t$&$\theta_{12}$&$\theta_{23}/10^{-2}$&$\theta_{13}/10^{-3}$&$\delta_{\rm CP}$&$J/10^{-5}$\\ \hline
 Reference&$6.15$&$1.35$&$2.68$&$3.11$ &$1.401$&$0.8616$&$0.22704$&$4.275$&$3.77$&$1.139$&$3.23$\\
 &$\pm 0.14$&$\pm 0.02$&$\pm 0.03$&$\pm 0.05$&$\pm 0.009$&$\pm 0.0043$&$\pm 0.00082$&$\pm 0.042$&$\pm 0.08$&$\pm 0.023$&$\pm 0.08$\\
 Sample&&&&&&&&&&&\\ \hline
 \#1& 6.1507 & 1.3491 & 2.6854 & 3.1097 & 1.4028 & 0.86196 & 0.22722 & 4.2634 & 3.7786 & 1.1328 & 3.1981\\
 \#2& 6.1822 & 1.3507 & 2.6809 & 3.1105 & 1.4016 & 0.86165 & 0.22741 &  4.2737 & 3.7671 & 1.1389 & 3.2075\\
 \#3& 6.1562 & 1.3495 & 2.6770 & 3.1113 & 1.4018 & 0.86172 & 0.22693 &  4.2727 & 3.7704 & 1.1392 & 3.2037\\
 \#4& 6.1563 & 1.3480 & 2.6802 & 3.1144 & 1.4008 & 0.86162 & 0.22719 &  4.2752 & 3.7703 & 1.1367 & 3.2051\\
 \#5& 6.1523 & 1.3505 & 2.6808 & 3.1128 & 1.4013 & 0.86159 & 0.22673 & 4.2727 & 3.7652 & 1.1420 & 3.2009\\
\end{tabular}
\end{ruledtabular}
\end{table*}

\subsection{\label{sec:constraints}Reference data}

The Standard Model quark sector has 10 observable quantities: the six Yukawa eigenvalues, $y_u$, $y_d$, $y_c$, $y_s$, $y_t$, $y_b$, and the four Cabibbo-Kobayashi-Maskawa (CKM) parameters, $\theta_{12}$, $\theta_{23}$, $\theta_{13}$, $\delta_{CP}$.
As experimental values of these quantities we use the $\overline{\rm MS}$ values at 1 TeV \cite{Antusch:2025fpm}, 
\begin{alignat}{2}\label{eqn:Antusch}
	y_u &= (6.15\pm 0.14) \times 10^{-6}, &\quad
	y_d &= (1.35\pm 0.02) \times 10^{-5},\crcr
	y_s &= (2.68\pm 0.03) \times 10^{-4}, &\quad
	y_c &= (3.11\pm 0.05) \times 10^{-3},\crcr
	y_b &= (1.401\pm 0.009) \time 10^{-2},&\quad
	y_t &= 0.8616\pm 0.0043,\crcr
	\theta_{12} &= 0.22704\pm 0.00082,&\quad
	\theta_{23} &= (4.275\pm 0.042)\times 10^{-2},\crcr
	\theta_{13} &= (3.77\pm 0.08)\times 10^{-3},&\quad
	\delta_{CP} &= 1.139\pm 0.023.
\end{alignat}
It is often convenient to use the Jarlskog invariant instead of $\delta_{CP}$, which is constrained to be \cite{Antusch:2025fpm}
\begin{align}
	J = 3.23\pm 0.08.
\end{align}
We investigate whether our flavour sector models can accommodate these values through adjustment of parameters $x_1,x_2,\ldots,x_{12}$.

In constructing the Yukawa couplings $y^u_{ij}$ and $y^d_{ij}$, the wave function of each chiral fermion is assumed to be localised at a point in the extra-dimensional space. 
By contrast, the Higgs-doublet wave function is taken to be flat over the region occupied by the chiral fermions, such that the Higgs vacuum expectation value is effectively constant and equal to
\begin{align}
	v = 246\text{ GeV}.
\end{align}
Then the mass matrices are defined in the usual way with the effective Yukawa matrices 
\begin{align}
	(M_u)_{ij}=\frac{v}{\sqrt 2}\, y^u_{ij},\qquad
	(M_d)_{ij}=\frac{v}{\sqrt 2}\, y^d_{ij},
\end{align}
which are diagonalised by pairs of unitary matrices as
\begin{align}
	D_u = V_d^\dag M_u U_u,\qquad
	D_d = V_u^\dag M_d U_d.
\end{align}
Numerically diagonalising the Hermitian matrices $M_uM_u^\dag$ and $M_dM_d^\dag$ one obtains the eigenvectors $V_u$, $V_d$ and quark masses
\begin{align}
	\begin{pmatrix}
		m_u^2 & &\\
		& m_c^2 &\\
		& & m_t^2
	\end{pmatrix}
	=D_uD_u^\dag = V_u^\dag M_uM_u^\dag V_u,\crcr
	\begin{pmatrix}
		m_d^2 & &\\
		& m_s^2 &\\
		& & m_b^2
	\end{pmatrix}
	=D_dD_d^\dag = V_d^\dag M_dM_d^\dag V_d.
\end{align}
The Yukawa eigenvalues are related to the quark masses 
\begin{align}
	m_a=\frac{v}{\sqrt 2}y_a, \quad a=u,d,c,s,t,b,
\end{align}
and the predictions for these quantities are compared with the constraints \eqref{eqn:Antusch}.
The CKM matrix is numerically computed as a product of the diagonalising matrices
\begin{align}\label{eqn:CKMdef}
	V_{\rm CKM}\equiv V_u^\dag V_d.
\end{align}
The CKM angles $\theta_{12}$, $\theta_{23}$, $\theta_{13}$ are found by comparing \eqref{eqn:CKMdef} with the standard (Chau-Keung) parametrisation
\begin{widetext}
\begin{align}\label{eqn:CKMstd}
	V_{\rm CKM}
	=
\left(\begin{array}{ccc}c_{12}c_{13} & s_{12}c_{13} & s_{13}e^{-i\delta_{CP}} \\-s_{12}c_{23}-c_{12}s_{23}s_{13}e^{i\delta_{CP}} & c_{12}c_{23}-s_{12}s_{23}s_{13}e^{i\delta_{CP}} & s_{23}c_{13} \\s_{12}s_{23}-c_{12}c_{23}s_{13}e^{i\delta_{CP}} & -c_{12}s_{23}-s_{12}c_{23}s_{13}e^{i\delta_{CP}} & c_{23}c_{13}\end{array}\right),
\end{align}
\end{widetext}
where $s_{ij}\equiv\sin\theta_{ij}$ and $c_{ij}\equiv\cos\theta_{ij}$.
In general the CKM matrix of the form \eqref{eqn:CKMdef} does not coincide with the parametrisation \eqref{eqn:CKMstd} due to phase rotation ambiguities, so we need to look at the absolute values of the components of \eqref{eqn:CKMdef} to determine the CKM angles $\theta_{12}$, $\theta_{23}$, and $\theta_{13}$.
The angle $\delta_{CP}$ is conveniently found from the Jarlskog invariant
\begin{align}\label{eqn:J}
	J&={\rm Im}\left[(V_{\rm CKM})_{12}(V_{\rm CKM})_{23}(V_{\rm CKM}^*)_{22}(V_{\rm CKM}^*)_{13}\right] \crcr
	&= s_{12}\,c_{12}\,s_{23}\,c_{23}\,s_{13}\,c_{13}^2\,\sin\delta_{CP}.
\end{align}
In the next two sections, we present the numerical methods used in our analysis and the resulting predictions of our models, examining their compatibility with the experimental constraints given in \eqref{eqn:Antusch}.

\begin{figure*}[t]
\includegraphics[width=175mm]{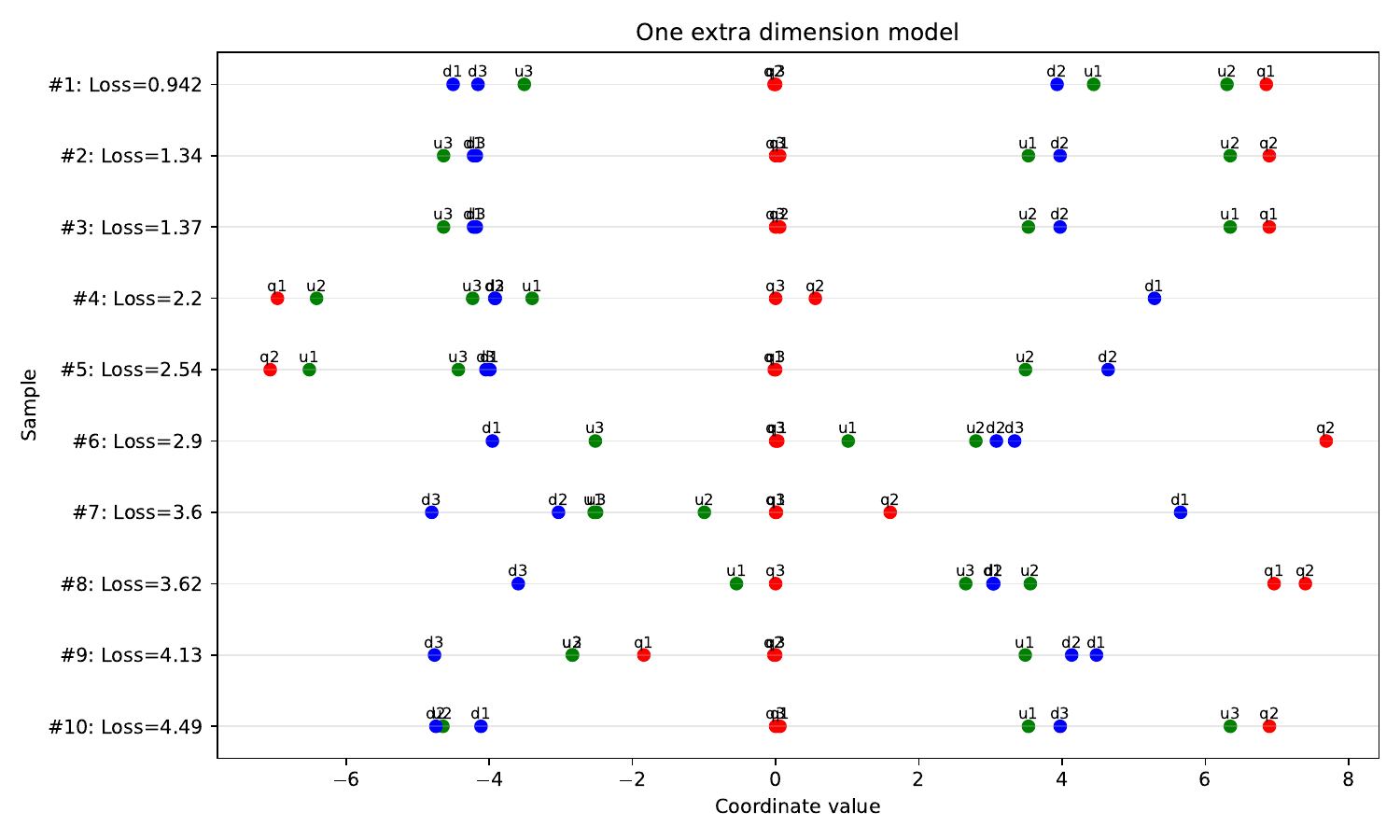}
\caption{\label{fig:config5d}
Field configurations of the 10 best-fit samples in the one-extra-dimension model. 
The left-handed quark fields $q_i$ are shown in red, while the right-handed up- and down-type quark fields, $u_i$ and $d_i$, are shown in green and blue, respectively. 
The field $q_3$ is fixed at the origin.
}
\end{figure*}

\section{\label{sec:5d}One extra dimension model}

\subsection{\label{sec:5d_methods}Numerical optimization}

Our goal is to determine the values of the model parameters $x_1,x_2,\ldots,x_{12}$ appearing in Eqs.~\eqref{eqn:y5} and \eqref{eqn:coord5} such that the model predictions reproduce the experimental values in \eqref{eqn:Antusch} as closely as possible. 
Since a grid scan of the 12-dimensional parameter space is computationally prohibitive, we employ optimisation techniques developed in machine learning that make use of algorithmic differentiation and random sampling. 
The key object in our analysis is a loss function that depends on the model parameters ${\mathbf x}=\{x_1,x_2,\ldots,x_{12}\}$.
We define it as
\begin{align}\label{eqn:L5}
	L_{\rm total}({\mathbf x})
	=L_{\rm qm}({\mathbf x})+L_{\rm CKM}({\mathbf x})+L_{\rm Jar}({\mathbf x})+L_{CP}({\mathbf x}).
\end{align}
The first term is the quark mass part, which reads
\begin{align}\label{eqn:Lqm}
	L_{\rm qm}({\mathbf x}) = \sum_a w_{\rm qm}^{(a)} \left|\ln\left|\frac{y_a(\mathbf x)}{y^{\rm cv}_a}\right|\right|,
\end{align}
where $a=u,d,c,s,t,b$ and $y_a({\mathbf x})$ are the prediction of the model.
The reference quantities $y^{\rm cv}_a$ are the centre values of \eqref{eqn:Antusch} and the coefficients $w_{\rm qm}^a$ are chosen to be the inverse of the $1\sigma$ relative error.
That is, for $y_a = y^{\rm cv}_a\pm \Delta y_a$ (e.g. $y^{\rm cv}_t = 0.8616$ and $\Delta y_t = 0.0043$), the coefficients are
\begin{align}
	w_{\rm qm}^{(a)}\equiv\frac{y^{\rm cv}_a}{\Delta y_a}.
\end{align}
With this choice, each term is calibrated so that $1\sigma$ deviation of $y_a({\mathbf x})$ from the centre value contributes unity to the total loss.
The second term is the CKM angle part, 
\begin{align}\label{eqn:LCKM}
	&L_{\rm CKM}({\mathbf x})\crcr
	&=w_{\rm CKM}^{(12)}\left|\ln\left|\frac{(V_{\rm CKM}({\mathbf x}))_{12}}{(V_{\rm CKM}^{\rm cv})_{12}}\right|\right| 
	+ w_{\rm CKM}^{(13)}\left|\ln\left|\frac{(V_{\rm CKM}({\mathbf x}))_{13}}{(V_{\rm CKM}^{\rm cv})_{13}}\right|\right|\crcr
	&\quad+ w_{\rm CKM}^{(23)}\left|\ln\left|\frac{(V_{\rm CKM}({\mathbf x}))_{23}}{(V_{\rm CKM}^{\rm cv})_{23}}\right|\right|.
\end{align}	
These three terms control the three CKM angles. 
Here, $V_{\rm CKM}({\mathbf x})_{ij}$ is the $(i,j)$ component of the predicted CKM matrix, $|(V_{\rm CKM}^{\rm cv})_{12}|=s_{12}c_{13}$, $|(V_{\rm CKM}^{\rm cv})_{13}|=s_{13}$, and $|(V_{\rm CKM}^{\rm cv})_{23}|=s_{23}c_{13}$ are given by the centre values of \eqref{eqn:Antusch}, and the coefficients are the inverse of the uncertainties, namely
\begin{align}
	\frac{1}{w_{\rm CKM}^{(12)}} &= \frac{\cos(\theta_{13}-\Delta\theta_{13})\sin(\theta_{12}+\Delta\theta_{12})}{\cos\theta_{13}\sin\theta_{12}} - 1,\crcr
	\frac{1}{w_{\rm CKM}^{(23)}} &= \frac{\cos(\theta_{13}-\Delta\theta_{13})\sin(\theta_{23}+\Delta\theta_{23})}{\cos\theta_{13}\sin\theta_{23}} - 1,\crcr
	\frac{1}{w_{\rm CKM}^{(13)}} &= \frac{\sin(\theta_{13}+\Delta\theta_{13})}{\sin\theta_{13}} - 1,
\end{align}
so that each $1\sigma$ deviation contributes one unit of loss.
The third term of \eqref{eqn:L5} controls the CP angle of the CKM matrix through the Jarlskog invariant,
\begin{align}\label{eqn:LJar}
	L_{\rm Jar}({\mathbf x}) = 
	w_{\rm Jar}\left|\ln\left|\frac{J({\mathbf x})}{J^{\rm cv}}\right|\right|,
\end{align}
where $J({\mathbf x})$ is the model prediction \eqref{eqn:J} , $J^{\rm cv}$ is the center value of \eqref{eqn:Antusch} and $w_{\rm Jar}=J^{\rm cv}/\Delta J=3.23/0.08$.
The last term is introduce to suppress CP to mitigate the strong CP problem,
\begin{align}\label{eqn:LCP}
	L_{CP}({\mathbf x}) = w_{CP}\left|\arg\det M_d({\mathbf x})\right|.
\end{align}
Although constraints on QCD CP violation, particularly those from the neutron electric dipole moment, are extremely stringent, we adopt a moderately large weight $w_{CP}=10^3$ in the optimisation analysis to maintain numerical tractability.

To numerically optimise the parameters $x_1, x_2, \ldots$, $x_{12}$, we generated 4096 random initial configurations from a uniform distribution over $-\pi \leq x_1, x_2, \ldots, x_{12} < \pi$. 
This range is natural for $x_1, \ldots, x_4$, which are phase parameters, and provides convenient ${\cal O}(1)$ bounds for $x_5, \ldots, x_{12}$.
The optimisation was performed using the vanilla Adam algorithm \cite{Kingma:2014vow} with its default hyperparameters, $\alpha = 0.001$, $\beta_1 = 0.9$, $\beta_2 = 0.999$, and $\epsilon = 10^{-8}$.

\begin{table*}[t]
\caption{\label{tab:table_6dparams}
Optimised parameter values $x_1, x_2, \ldots, x_{12}$ for the 5 best fit samples for the two extra dimension model.
The optimised loss function values for these 5 samples are 0.457,
 0.462,
 0.482,
 0.486, and
 0.499.}
\begin{ruledtabular}
\begin{tabular}{ccccccccccccc}
 Sample &\multicolumn{12}{c}{Optimised parameter values}\\
(type) &$x_1$&$x_2$&$x_3$&$x_4$&$x_5$&$x_6$&$x_7$&$x_8$&$x_9$&$x_{10}$&$x_{11}$&$x_{12}$\\ \hline
 \#1 (d)& -1.4574 & 1.2402 & 4.3009 & -2.1939 & -0.043863 & 0.97873 & 2.7282 & 0.0056122 & -1.9345 & -3.8797 & -2.8570 & -1.2032 \\
 \#2 (7)&  2.1362 & 1.7137 & 2.6881 & -2.2426 & 0.034977 & 0.99208 & 2.7684 & -0.0086459 & -1.9004 & -3.8944 & -2.8538 & -1.1421 \\
 \#3 (d)& -0.43696 & 2.4282 & 0.83581 & -2.5946 & -0.10209 & -0.87904 & 3.0463 & 0.21507 & -3.2692 & 4.3672 & -2.7787 & -1.6416 \\
 \#4 (5)& 1.1424 & 1.2041 & 1.2432 & -2.0543 & -0.0062153 & 0.94375 & 2.5559 & 0.24881 & -2.0393 & -3.9090 & -2.8006 & -1.1090 \\
 \#5 (7)& -0.66948 & 2.3715 & 0.70856 & -3.0679 & 0.15047 & -0.83097 & 2.9951 & -0.55299 & -4.0532 & -4.3160 & 2.9002 & -1.8187 \\ 
\end{tabular}
\end{ruledtabular}
\end{table*}

\begin{figure}[b]
\includegraphics[width=95mm]{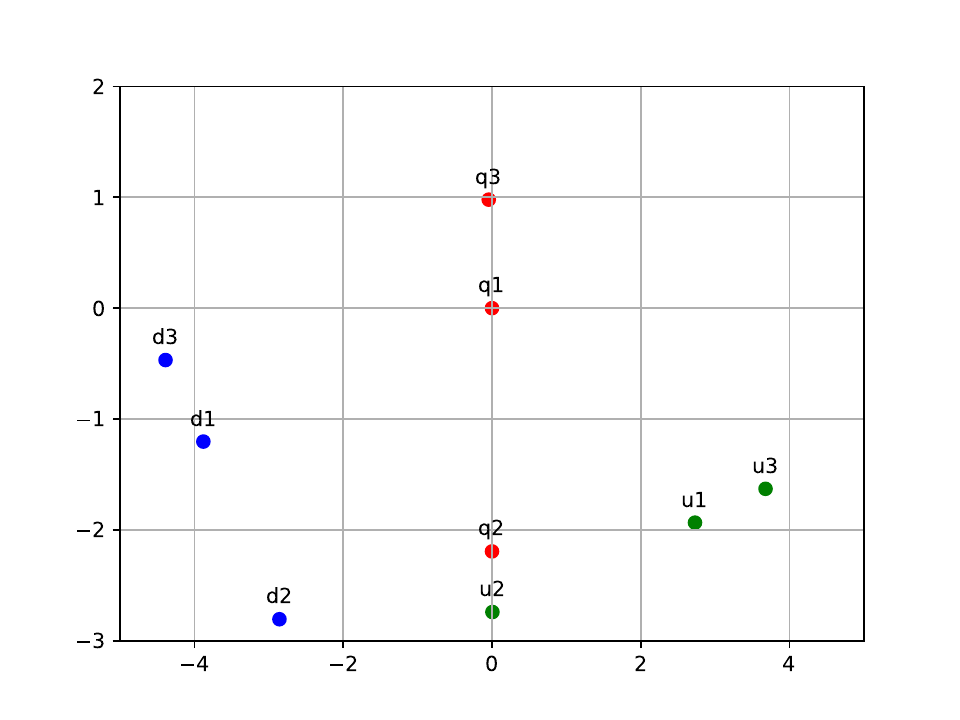}
\caption{\label{fig:config6d_best} 
Optimised field configuration of the best performing sample in the two extra dimension model.
The left-handed quark fields $q_i$ are shown in red, while the right-handed up- and down-type quark fields, $u_i$ and $d_i$, are shown in green and blue, respectively. 
The field $q_1$ is fixed at the origin and the field $q_2$ is placed on the vertical axis.
}
\end{figure}

\subsection{\label{sec:5d_results}Results}

We performed numerical optimisation on 4096 samples for $10^5$ iterations using the Adam optimiser \cite{Kingma:2014vow} to minimise the loss function. 
Since the loss function for the Jarlskog invariant, Eq.~\eqref{eqn:LJar}, is insensitive to the sign of $J$, some samples converged to solutions with negative values of $J$; these samples were excluded from the analysis. 
Table~\ref{tab:table_5dparams} lists the optimised parameter values for the five best-fit samples. 
The optimised value of the total loss function for Sample \#1, which provides the best fit, is 0.942.
Table~\ref{tab:table_5dfit} shows the predictions obtained from the five samples listed in Table~\ref{tab:table_5dparams}, together with the reference values used in the optimisation. 
The loss function is normalised such that a deviation of one standard deviation in a Standard Model observable contributes unity. 
Since constraints are imposed on six quark masses and four CKM parameters, the loss function is expected to be $\sim 10$ if each quantity deviates from its central value by $1\sigma$. 
The optimised loss values of $\sim 1$ therefore indicate an excellent fit.
Given that the model contains 12 free parameters and is required to satisfy 10 constraints (or 11 when the CP-violation loss term is included), the existence of solutions is expected. 
Nevertheless, finding such solutions is a nontrivial task and our results demonstrate that a phenomenologically viable fit can indeed be achieved by the method of numerical optimisation.

Figure~\ref{fig:config5d} shows the field configurations of the 10 best-fit samples, together with the corresponding optimised values of the loss function shown on the left. 
This figure highlights several features that are not immediately apparent from Table~\ref{tab:table_5dparams}. 
In particular, the generation labels assigned to fields of the same type (e.g. $u_1$ and $u_2$) are not physically meaningful, since the observables are ultimately determined by the eigenvalues and eigenvectors. 
Consequently, Samples \#2 and \#3 in Fig.~\ref{fig:config5d} should be regarded as representing the same configuration. 
Although not all optimised samples converge to a unique configuration, Fig.~\ref{fig:config5d} reveals several common features. 
In all cases, there is a pair of $q$ (red) and $u$ (green) fields located in close proximity, accounting for the ${\cal O}(1)$ top Yukawa coupling. 
In many of the samples (with the exception of \#4 and \#8), two $q$ fields also tend to cluster closely together. 
The underlying reason for this behaviour is not presently clear to the authors.

\begin{table*}[t]
\caption{\label{tab:table_6dfit}
Predictions for the Standard Model parameters obtained from the five best-fit samples listed in Table~\ref{tab:table_6dparams}. 
The reference experimental values are the $\overline{\rm MS}$ quantities evaluated at 1 TeV \cite{Antusch:2025fpm}. 
The $1\sigma$ uncertainties are used to define the weights in the loss function.
}
\begin{ruledtabular}
\begin{tabular}{cccccccccccc}
 &\multicolumn{11}{c}{Standard Model parameters}\\
 &$y_u/10^{-6}$&$y_d/10^{-5}$&$y_s/10^{-4}$&$y_c/10^{-3}$&$y_b/10^{-2}$&$y_t$&$\theta_{12}$&$\theta_{23}/10^{-2}$&$\theta_{13}/10^{-3}$&$\delta_{\rm CP}$&$J/10^{-5}$\\ \hline
 Reference&$6.15$&$1.35$&$2.68$&$3.11$ &$1.401$&$0.8616$&$0.22704$&$4.275$&$3.77$&$1.139$&$3.23$\\
 &$\pm 0.14$&$\pm 0.02$&$\pm 0.03$&$\pm 0.05$&$\pm 0.009$&$\pm 0.0043$&$\pm 0.00082$&$\pm 0.042$&$\pm 0.08$&$\pm 0.023$&$\pm 0.08$\\
 Sample &&&&&&&&&&&\\ \hline
 \#1 & 6.1634 & 1.3474 & 2.6793 & 3.1125 & 1.4014 & 0.86186 & 0.22679 & 4.2766 & 3.7698 & 1.1394 & 3.2045 \\
 \#2 & 6.1785 & 1.3493 & 2.6805 & 3.1128 & 1.4008 & 0.86157 & 0.22701 & 4.2756 & 3.7717 & 1.1394 & 3.2083 \\
 \#3 & 6.1542 & 1.3507 & 2.6821 & 3.1125 & 1.4019 & 0.86169 & 0.22681 & 4.2776 & 3.7672 & 1.1418 & 3.2070 \\
 \#4 & 6.1407 & 1.3495 & 2.6802 & 3.1086 & 1.4019 & 0.86175 & 0.22701 & 4.2751 & 3.7687 & 1.1393 & 3.2052 \\
 \#5 & 6.1481 & 1.3494 & 2.6781 & 3.1076 & 1.4000 & 0.86174 & 0.22715 & 4.2723 & 3.7732 & 1.1385 & 3.2078 \\
\end{tabular}
\end{ruledtabular}
\end{table*}

\begin{figure}[b]
\includegraphics[width=90mm]{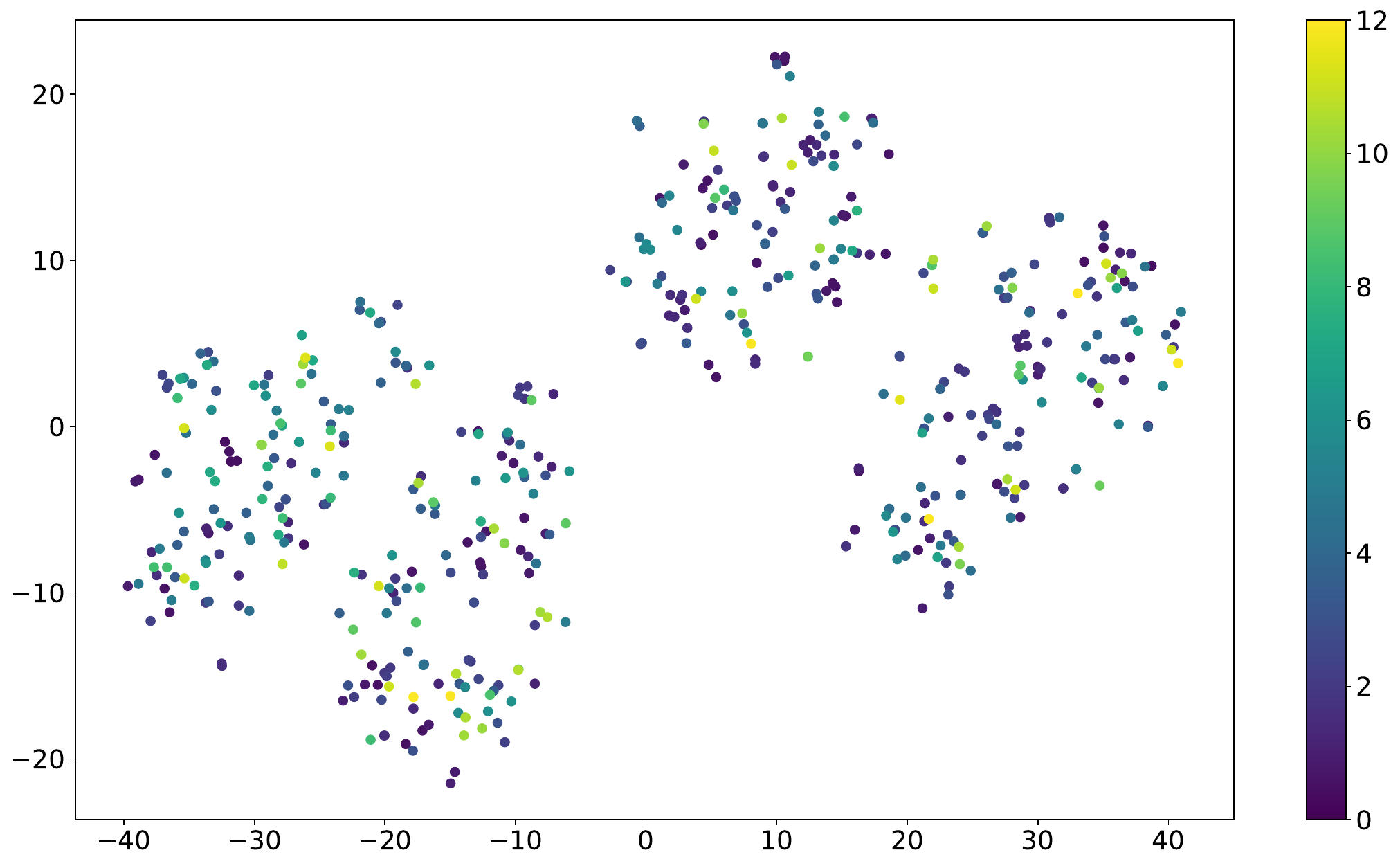}
\caption{\label{fig:tSNE} 
Planar visualisation of 12 dimension parameters $\{x_1,x_2,\ldots,x_{12}\}$ of the two extra dimension model, using dimensionality reduction algorithm of t-SNE.
The plot shows 502 best-fit samples with optimised loss values $<12$.
The colour represents the values of the optimised loss function: darker blue for smaller loss values, indicating better performance. 
}
\end{figure}

\begin{figure*}[t]
\includegraphics[width=175mm]{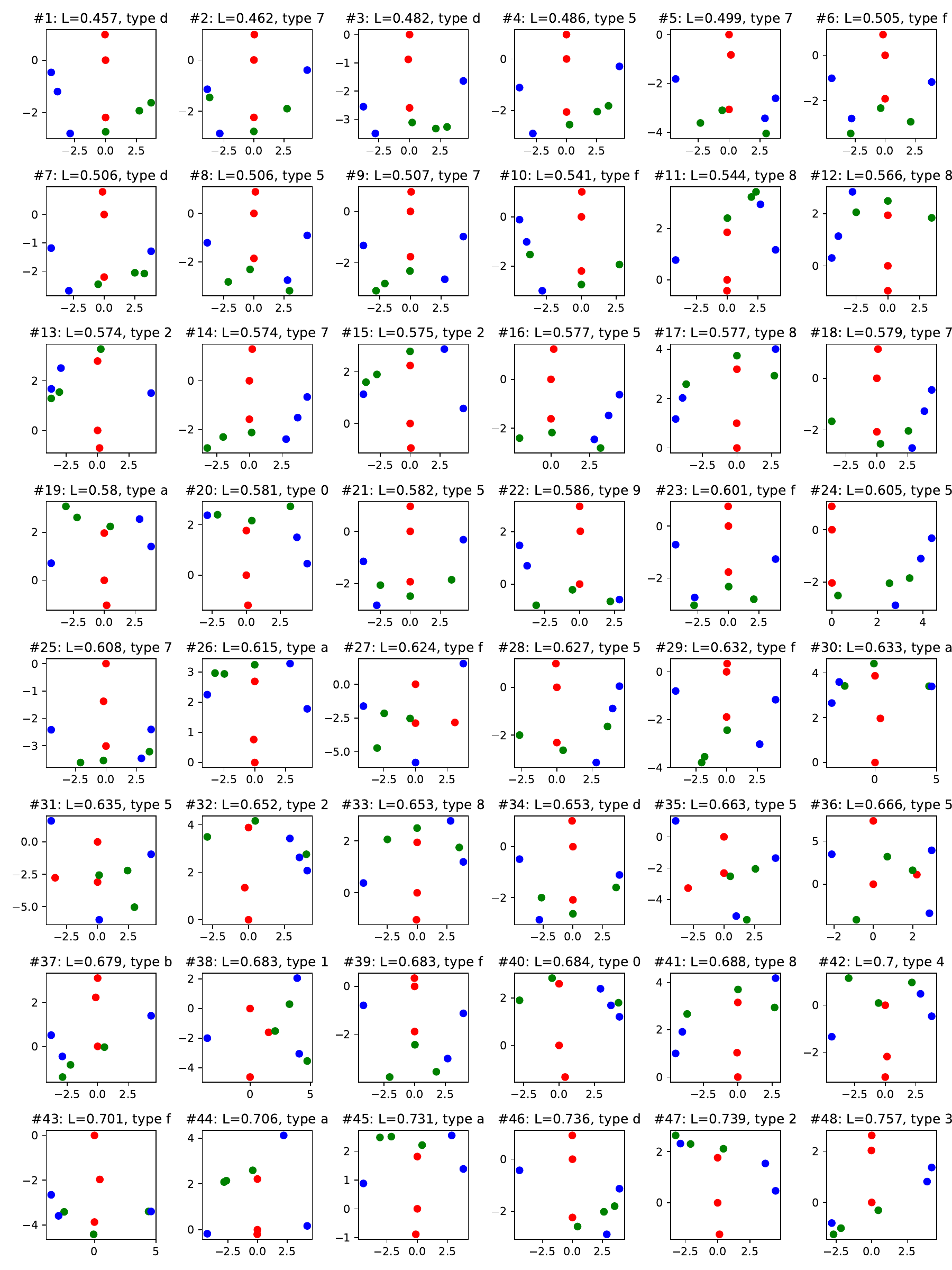}
\caption{\label{fig:config6d_48} 
Optimised field configuration of the 48 best-fit samples in the two extra dimension model.
The left-handed quark fields $q_i$ are shown in red, while the right-handed up- and down-type quark fields, $u_i$ and $d_i$, are shown in green and blue, respectively. 
The field $q_1$ is fixed at the origin and the field $q_2$ is constrained on the vertical axis.
}
\end{figure*}

\section{\label{sec:6d}Two extra dimension model}

\subsection{\label{sec:6d_methods}Numerical optimisation}

The analysis of the two extra dimension model parallels that of the previous section, except for a couple of changes in technicalities. 
Our goal is again to find the values of model parameters $x_1,x_2,\ldots,x_{12}$ that satisfy the experimental constraints \eqref{eqn:Antusch}, and we use numerical optimisation of a loss function to achieve this goal.
Since the mass matrices $M_u$ and $M_d$ are Hermitian by construction, the CP is automatically conserved in this case. 
We thus consider a loss function
\begin{align}\label{eqn:L6}
	L_{\rm total}({\mathbf x})
	=L_{\rm qm}({\mathbf x})+L_{\rm CKM}({\mathbf x})+L_{\rm Jar}({\mathbf x}),
\end{align}
where the three terms are defined by \eqref{eqn:LCKM}, \eqref{eqn:Lqm}, and \eqref{eqn:LJar}.
As discussed in the preceding section, each term is normalised by the precision of the experimental data so that $1\sigma$ deviation corresponds to unity of loss.

Solving the Hermiticity conditions yields $2^4=16$ distinct cases, corresponding to the independent sign choices in Eqs.~\eqref{eqn:w2}, \eqref{eqn:z3}, \eqref{eqn:w2t}, and \eqref{eqn:z3t}. 
We refer to this classification as the solution {\em type} and label the 16 possibilities by the hexadecimal digits $0,1,\ldots,e,f$. 
Since the different types are independent, we performed parameter optimisation separately for each type, starting from 4096 randomly generated initial configurations.
The square roots appearing in Eqs.~\eqref{eqn:w2}, \eqref{eqn:z3}, \eqref{eqn:w2t}, and \eqref{eqn:z3t} complicate the optimisation procedure, since it cannot be continued once any parameter becomes complex. 
The random initial configurations are therefore chosen from
$-\pi \leq x_1, x_2, \ldots, x_7, x_9, x_{10}, x_{12} < \pi$ and 
\begin{align*}
	&-\sqrt{(x_4-x_9)^2+x_7^2}<x_8<\sqrt{(x_4-x_9)^2+x_7^2},\crcr
	&-\sqrt{(x_4-x_{12})^2+x_{10}^2}< x_{11} < \sqrt{(x_4-x_{12})^2+x_{10}^2}
\end{align*}
so that the arguments of the square roots in Eqs.~\eqref{eqn:w2} and \eqref{eqn:w2t} remain non-negative. 
Initial samples for which Eqs.~\eqref{eqn:z3} or \eqref{eqn:z3t} yield complex values are simply discarded. 
During the optimisation process, some samples also evolve into regions of parameter space where complex values arise; such samples are likewise excluded from the analysis.
We use the standard Adam optimiser \cite{Kingma:2014vow} with hyperparameters 
$\alpha = 0.001$, $\beta_1 = 0.9$, $\beta_2 = 0.999$, and $\epsilon = 10^{-8}$.

\subsection{\label{sec:6d_results}Results}

We performed numerical optimisation for $2\times 10^5$ iterations on 4096 samples, for each of the 16 solution types.
Some samples developed complex parameter values as a result of the Hermiticity constraints, Eqs.~\eqref{eqn:w2}, \eqref{eqn:z3}, \eqref{eqn:w2t}, and \eqref{eqn:z3t}, and were excluded from the analysis. 
Samples yielding negative values of the Jarlskog invariant were likewise discarded. 
After applying these selection criteria, a total of 4412 samples remained.

Table~\ref{tab:table_6dparams} lists the optimised parameter values of the five best performing samples. 
The best fit sample (\#1) is type d (the signs of \eqref{eqn:w2}, \eqref{eqn:z3}, \eqref{eqn:w2t}, \eqref{eqn:z3t} are $--+-$ in this order) and its loss function reaches 0.457.
The field configuration of this sample is shown in Fig.~\ref{fig:config6d_best}. 
The predictions for the Standard Model parameters by these five samples are listed in Table~\ref{tab:table_6dfit}. 
As the loss function \eqref{eqn:L6} is normalised such that a deviation of one standard deviation in each of the 10 Standard Model observables contributes unity to the total loss, optimised loss values below 0.5 indicate an excellent fit. 
This problem involves optimising 12 parameters subject to 10 constraints, which is mathematically feasible. 
The present analysis demonstrates that this objective can be successfully achieved using our methods.

Figure \ref{fig:config6d_48} shows the field configurations of the 48 best-performing samples out of 4412. 
The optimised values of the loss function and the sample types, classified according to the sign choices of the constraints \eqref{eqn:w2}, \eqref{eqn:z3}, \eqref{eqn:w2t}, and \eqref{eqn:z3t}, are also indicated. 
As in the model with one extra dimension, the visualisation reveals several characteristic features of the optimised parameter configurations. 
We omit the generation labels of the fields, since the ordering of the generations prior to diagonalisation has no physical significance. 
A pair of $q$ (red) and $u$ (green) fields can often be seen in close proximity, thereby realising an ${\cal O}(1)$ top Yukawa coupling. 
Although the field $q_1$ is fixed at the origin and $q_2$ is constrained to lie on the vertical axis by construction, the third left-handed quark field, $q_3$, also tends to align with the vertical axis in many, though not all, samples. 
In such cases, some configurations, such as \#1 and \#2, are related to one another by reflection of $u$ and/or $d$ fields about the vertical axis. 
All samples shown in Fig.~\ref{fig:config6d_48} achieve optimised loss values below 1 and are in excellent agreement with the experimental constraints.

As we have normalised the loss function such that each $1\sigma$ deviation contributes unity to the total loss, and there are 10 constraints (six from the quark masses and four from the CKM parameters), an optimised loss value of $\lesssim {\cal O}(10)$ is adopted as the criterion for a viable sample. 
In our analysis, 502 samples out of 4412 achieved optimised loss values below 12. 
These 502 best-performing samples are displayed in Fig.~\ref{fig:tSNE} using a planar representation obtained with the t-distributed stochastic neighbour embedding (t-SNE) algorithm, where darker blue (brighter yellow) colours correspond to smaller (larger) optimised loss values. 
Several clusters can be identified; however, all clusters appear to be of comparable fit quality, as the colour distribution exhibits no obvious bias. 
One might expect the landscape of viable parameter configurations in the 12-dimensional parameter space to possess some characteristic structure, but its detailed features remain unclear at present.

\section{\label{sec:final}Final remarks}

In this paper, we constructed the quark Yukawa couplings of the Standard Model as overlap integrals of wave functions in extra dimensions and investigated whether they are compatible with experimental constraints. 
We examined two concrete models: the first with one extra dimension and the second with two extra dimensions, in which Hermiticity conditions are explicitly imposed on the Yukawa matrices. 
Both models involve 12 free parameters, and we employed numerical optimisation techniques from machine learning to assess their viability. 
We defined a precision-weighted loss function such that each $1\sigma$ deviation from the corresponding central value contributes unity to the total loss. 
For both the one- and two-extra-dimensional models, we demonstrated that this method identifies multiple parameter configurations that provide excellent fits to the experimental data. 
Our construction therefore provides a systematic framework for generating mass matrices and may be regarded as an alternative to the Froggatt-Nielsen mechanism \cite{Froggatt:1978nt} or the democratic mass matrix model \cite{Harari:1978yi}. 
The resulting mass matrices may be characterised as possessing an extra-dimensional {\em geometric texture}, linking geometric field configurations in extra dimensions to the precision science of flavour physics.

Let us conclude with several remarks on our results. 
The Yukawa-coupling models considered in this work, Eqs.~\eqref{eqn:y5}, \eqref{eqn:yu6}, and \eqref{eqn:yd6}, are relatively simple and generic. 
In the one-extra-dimensional model, we assume the natural translational and phase-rotation symmetries of the extra dimension, while in the two-extra-dimensional model we further impose Hermiticity of the Yukawa matrices and rotational symmetry in the extra dimensions. 
The fact that these models yield numerous parameter configurations with excellent fits to the experimental data suggests that a substantial degree of flexibility remains. 
Consequently, one or two additional constraints could likely be imposed while still maintaining agreement with observations at the $2\!-\!3\sigma$ level. 
Reducing the number of tuneable parameters from 12 to 11 or 10 would restrict the space of viable configurations and may ultimately allow the structure of the quark flavour sector to be determined uniquely. 
The choice of additional constraints, however, depends on the specific theoretical framework in which the models are embedded. 
An obvious direction for future investigation is the interpretation of our results within concrete model-building frameworks, such as those motivated by string theory. 

Our analysis also reveals several characteristic features of the viable parameter configurations. 
In particular, Fig.~\ref{fig:config6d_48} suggests that the field configurations in the two-extra-dimensional model tend to exhibit a number of recurring geometric patterns, while Fig.~\ref{fig:tSNE} indicates the possible existence of a clustering structure among viable configurations in the 12-dimensional parameter space. 
Owing in part to the limitations of our computational resources, however, the details of this structure remain unclear, and a more comprehensive exploration of the parameter space is clearly desirable. 
Whether a meaningful classification of experimentally viable parameter configurations is possible, and, if so, how many distinct clusters may exist, are questions that warrant further investigation. 
Finally, although we have not considered the lepton sector in the present work, it can be constructed and analysed within the same framework, subject to assumptions regarding the neutrino sector, including the mass ordering and the mass of the lightest neutrino. 
Such an analysis is currently underway, and we hope to report the results in the near future.

\begin{acknowledgments}
This work was supported in part by the United States Department of Energy Grant 
Nos. DE-SC0012447, DE-SC0023713, and DE-SC0026347 (N.O.).
\end{acknowledgments}




%


\end{document}